\documentclass[%
 aip,
 jap,%
 amsmath,amssymb,
 preprint,%
 nofootinbib,
]{revtex4-2}

\usepackage{graphicx}
\usepackage{dcolumn}
\usepackage{bm}
\usepackage[mathlines]{lineno}
\usepackage{setspace}

\begin{document}


\title{Fe$^{2+}$:forsterite Qubit System}


\author{Dishaanth S. Pillai}
\email{dishaanth49@gmail.com}
\affiliation{Andover Central High School, Andover, KS, USA}

\author{G. M. Wysin}
\email{wysin@k-state.edu}
\affiliation{Department of Physics, Kansas State University, Manhattan, KS 66506-2601}%


\date{\today}

\begin{abstract}
Quantum Computing uses quantum mechanical phenomena, such as superposition and entanglement of quantum bits (qubits), to represent and manipulate information in multidimensional state spaces. This approach allows complex mathematical functions to be executed simultaneously rather than sequentially, making them much more efficient than classical computers. However, current quantum computers suffer from quantum decoherence, where small changes to their environment, such as thermal fluctuations and vibrations, affect the systems' fragile states. The goal of this study is to discuss the functionality of a proposed Fe$^{2+}$:forsterite qubit system by solving the unitary time-evolution of a quantum state vector for the system's Hamiltonian, using known axial and traverse anisotropy terms at temperatures up to 10 K. The system can effectively transfer population between its two lowest states at rates predicted by the Rabi half-period, with 99.9\% fidelity. Additionally, it can hold the population at a desired state without docohering. The results remained consistent for temperatures up to 10 K, meaning, unlike many common qubit systems, the proposed qubit system could function without a dilution fridge and is not fragile to thermal noise. 
\end{abstract}

\maketitle


\section{Introduction}
\label{sec:introduction}

Classical computers use classical bits that have binary values: 0 or 1, acting as a basic light switch, either on or off. Quantum computers use quantum bits (qubits) that can have values of 0, 1, or both simultaneously through superposition. They can entangle with other qubits, allowing them to process large numbers of possibilities at once, precisely $2^n$ possibilities, where $n$ represents the number of qubits. However, many current qubit systems are incredibly fragile, as they are subject to unwanted coupling to their environment, energy relaxation, and dephasing caused by thermal noise, electromagnetic interference, and material defects, leading to decoherence and high rates of calculation error \cite{Kjaergaard2020}. Despite this, quantum computers are a heavily-researched field in today's world, with many new models being developed every day to improve the current bottlenecks in quantum computing, such as the extreme cooling conditions needed for quantum computing platforms \cite{Lucas2023}; the material engineering complexity for creating degenerate states\cite{KramersSMMSwitching}; and the difficulty in driving transitions across large angular momentum gaps\cite{KramersSMMSwitching}. 

There have been numerous proposed qubit systems, each having different characteristics and behaviors. One such system is a single non-Kramers ion, a paramagnetic ion with an even number of electrons in its unfilled outer shell, with an integer spin number (e.g. $S = 1, 2$), and $2S+1$ associated states $|m\rangle$, labeled by a magnetic quantum number $m$ ranging from $-S$ to $+S$. For a Kramers ion, Kramers' theorem\cite{Kramers1930} states that with an odd number of unpaired electrons, every energy level must remain at least doubly degenerate unless time-reversal symmetry is broken. For a non-Kramers ion, with an even number of unpaired electrons, there is no such guarantee of double degeneracy\cite{KramersSMMSwitching}; the crystal-field alone is enough to completely lift degeneracy, splitting all $2S+1$ sublevels apart. This splitting is influenced by two anisotropy parameters: axial ($D$), which sets an energy preference along the crystal's easy axis and separates states of different $|m|$ in energy, and transverse ($E$), which breaks the axial symmetry and mixes states differing by $\Delta m = \pm 2$. If the crystal has the appropriate $D$ and $E$, the transverse mixing of states produces magnetization tunneling\cite{ChudnovskyGunther1988} that connects the two lowest-energy ($m \approx \pm 2$) states virtually through the $m = 0$ level. This produces a small, controlled tunnel splitting between the two lowest-energy states. The splitting remains small because connecting the two lowest-energy states requires an indirect, second-order process virtually through $m = 0$, which is inherently weaker than the direct, first-order coupling that connects two adjacent $m$ states under ordinary anisotropy. One tradeoff with non-Kramers ions is that they are more susceptible to decoherence from crystal-field fluctuations. 

Using the two lowest-energy states of a non-Kramers ion to represent the states of a qubit has been done before by Smoleński et al. \cite{Smolenski2016}, who subjected a Fe$^{2+}$ ion embedded in CdSe/ZnSe(as a quantum dot) to a high strain, arising from lattice-mismatched epitaxial growth, to change the ion's energy level structure.  This produced a nearly doubly degenerate ground state with $m =\pm 2$. 
They detected a magnetic character optically through photoluminescence measurements, exploiting the coupling between a confined exciton and the single iron impurity's spin state. This degeneracy in their system only exists because of deliberately engineered strain from the epitaxial growth process and is not a natural property of the crystal. They also only ran a spin-state characterization and a readout demonstration without driving the system or discussing any controlled qubit operations. To our knowledge, there hasn't been a proposed material system with $m = \pm 2$ where splitting arises naturally from a bulk crystal's own intrinsic structure and includes an actual proposed mechanism for coherent gate control. 

We propose a single Fe$^{2+}$ ion substituting for Mg$^{2+}$ in forsterite (Mg$_2$SiO$_4$), specifically in the M2 crystallographic site, as a functional qubit operating between the two lowest-energy states, with reliable behavior. The system is studied to determine the stability and relaxation properties of the states, using Hamiltonian diagonalization and time-dependent master equation simulations. We use the eigenstates of the Hamiltonian matrix to identify the energy levels and implement a strain/sound acoustic wave (SAW), oscillating the crystal-field transverse anisotropy ($E$) to coherently drive the transition between the two states. To confirm the accuracy of our method, we compare its predicted energy levels against three independent benchmark systems with known results, including an independent implementation of the same Hamiltonian using Qiskit \cite{Qiskit}, an open-source software framework for building and simulating quantum circuits. We characterize the decoherence and temperature behavior of the model; the system behaves robustly even at 10 K. We discuss two possible methods of readout for this system. 

\section{System and model}
\label{sec:system_model}

Our proposed model includes an iron ion in the +2 oxidation state  (Fe$^{2+}$), with 6 electrons in its 3d subshell. In an isolated ion, these five orbitals are degenerate, and Hund's rule states that the electrons will try to fill each orbital individually before pairing up, maximizing spin. However, once substituted into forsterite, Fe$^{2+}$ sits at the M2 site, coordinated by neighboring oxygen ions, whose surrounding ligand field breaks the degeneracy. The five 3d orbitals split into two energy groups, separated by a crystal-field splitting energy arising from the crystal-field anisotropies. 

There are two factors affecting how the 6 electrons are split: Hund's rule favoring the electrons occupying degenerate orbitals singly before pairing, giving the unpaired electrons a high-spin configuration; and the crystal-field splitting energy favoring the pairing of electrons in the lower orbital set, resulting in a low-spin configuration. A weak-field ligand environment would cause small splitting, giving the 4 unpaired electrons a high-spin configuration, constituting a total spin angular momentum of $S = 2$. A strong-field ligand environment causes large splitting, meaning that all electrons are paired, constituting a total spin angular momentum of $S = 0$. Oxygen is a weak-field ligand, so Fe$^{2+}$ is expected to have a total spin angular momentum of $S = 2$, as verified by Kozanecki et al \cite{Kozanecki2017}. This is critical for the proposed qubit system because a nonzero total spin ($S = 2$) gives Fe$^{2+}$:forsterite multiple distinct sublevels to encode two quantum states; however, if forsterite had been in a strong-field ligand environment, then $S$ would equal 0, leaving no sublevels to encode quantum states, making a qubit system impossible. 

In the crystal structure, magnesium ions occupy two crystallographically distinct octahedral sites, conventionally named M1 and M2. These two sites cannot be treated the same, as they have their own distinct symmetry and geometry; thus, they have different axial and transverse anisotropy. Fe$^{2+}$ substitutes for Mg$^{2+}$ at approximately a 1:1 ratio among the M1 and M2 sites \cite{McCarty2015}. This study focuses on the M2 site because crystal-field parameters ($D$ and $E$) are available from Kozanecki et al. \cite{Kozanecki2017} for this specific site. Nevertheless, the  M1 site is still a feasible alternative and could be modeled as a qubit system in future works if the equivalent $D$ and $E$ values are measured. 

The system was modeled using a spin Hamiltonian with two parts: a zero-field splitting (ZFS) term, arising from dipolar interactions within the crystal, and a Zeeman term describing the effect of an applied field:
\begin{equation}
    \hat{H} = D\hat{S}_z^2 + E(\hat{S}_x^2 - \hat{S}_y^2) + g_L \mu_B B_z \hat{S}_z .
    \label{eq:hamiltonian_static}
\end{equation}
This is a standard equation used to describe how a crystal field acts on an ion's spin degrees of freedom. The first two terms describe the zero-field splitting arising from the crystal's axial and transverse anisotropy. The values of $D$ and $E$ were obtained by Kozanecki et al. \cite{Kozanecki2017}, specifically from their S2b scheme's M2 site. $\hat{S}_x$, $\hat{S}_y$, and $\hat{S}_z$ are the spin operators for spin $S=2$. The third term describes the effect of an applied magnetic field. $g_{L} \approx 2$ is the Landé $g$-factor or gyromagnetic ratio factor, which shows the strength of a particle's magnetic moment relative to its spin. $\mu_{B}=e \hbar /\left(2 m_{\mathrm{e}}\right)$ is the Bohr magneton, which is the natural unit of measurement for a single electron's magnetic dipole. We diagonalize the Hamiltonian in order to obtain the eigenvalues and eigenstates that define the qubit system's stationary states. Throughout the study, energies are represented in their equivalent temperatures in Kelvin; all literature values reported in other units are converted accordingly. 

Shakurov et al. \cite{Shakurov2010} originally performed an electron paramagnetic resonance measurement on Fe$^{2+}$ in forsterite; however, they assumed a total spin angular momentum of $S' = 1$, which is an inaccurate simplification, given that the established value for the Fe$^{2+}$ ion is $S = 2$. Kozanecki et al. \cite{Kozanecki2017} published a paper that converted these fictitious $S' = 1$ values to the accurate $S = 2$ values. This conversion resulted in multiple schemes of $D/E$ values (S2a, S2b, S2b', S2c). The paper directly rules out both the S2a and S2c; S2a for it's rather unlikely sign for the $D$ value and S2c because it's experimental temperatures (4.2-15 K) sit at a population state that is unfavorable according to Boltzmann statistics. S2b$'$'s $E$/$D$ ratio sits on the boundary of the conventional range (1/3), while S2b's $E$/$D$ ratio sits comfortably within the conventional range. As a result, this study uses the values from the S2b scheme: $D \approx -12.37$\,K and $E \approx 0.896$\,K. 

Diagonalizing the Hamiltonian matrix analytically in Appendix \ref{app:S=2_analytic_solution} at zero applied field $B_z$ results in five energy eigenstates. They are summarized in Eq.\ (\ref{S2evs}). Because the transverse term $E(\hat{S}_x^2 - \hat{S}_y^2)$ couples states differing by $\Delta m = \pm 2$, the Hamiltonian is not diagonal in the $\hat{S}_z$ eigenbasis $|m\rangle$. Instead, it separates into two independent blocks: \{$m = -2, 0, +2$\} and \{$m = -1, +1$\}. The five zero-field energy eigenvalues of the system are $-49.67$ K, $-49.48$ K, $-15.06$ K, $-9.68$ K, and $0.19$ K. The lowest and next lowest energy states are the symmetric (with a small $m = 0$ admixture) and antisymmetric superpositions of $|+2\rangle$ and $|-2\rangle$, respectively, forming a nearly degenerate ground-doublet. The highest-energy state, denoted $|\psi_{20}^{+}\rangle$, is dominated by $m = 0$ and lies 49.86 K above the lowest state ($0.19 - (-49.67)$ K). As a result of the $E$-term only connecting $\Delta m = \pm 2$, the $m = \pm 2$ states only couple to each other virtually through $m = 0$. Only the symmetric combination can mix with $m =$ 0, which lowers its energy, while the antisymmetric combination is unaffected and remains at $4 D$ = $-49.48$ K. The resulting tunnel splitting is $\Delta =$ 0.19 K, the difference between the two lowest eigenvalues, which corresponds to a frequency of 4.040 GHz [Eq.\ (\ref{tunnelf})]. The mixing is weak and perturbative: the unperturbed separation between $m = \pm 2$ and $m = 0$ ($4|D| \approx 49.48$\,K) is much larger than the transverse coupling that connects them to $m = 0$ ($\sqrt{6}E$ $\approx$ 2.2\,K), producing a splitting ($\Delta \approx$ 0.19 K) that is itself much smaller than the coupling. 

For the discussion of qubit operations, we define the two qubit basis states $|\mathbf{0}\rangle$ and $|\mathbf{1}\rangle$ as the nearly degenerate ground-doublet eigenstates dominated by $m = \pm 2$. Bold font is used to distinguish them from the eigenstates $|m\rangle$ of $\hat{S}_z$. In the notation in Appendix \ref{app:S=2_analytic_solution}, Eq.\ (\ref{S2evs}), they are 
\begin{align}
     |\mathbf{1}\rangle &\equiv |\psi_2\rangle = \tfrac{1}{\sqrt{2}}\left(|2\rangle-|-2\rangle\right), \nonumber \\ |\mathbf{0}\rangle &\equiv |\psi_{20}^{-}\rangle \approx \tfrac{1}{\sqrt{2}} \left(|2\rangle+|-2\rangle\right).    
\end{align}

As an approximate two-level system, these are its zero-field stationary states when not disturbed by any perturbations, composed with equal probabilities of $m=+2$ and $m=-2$ . Alternatively, if the system starts exclusively in $|m=2\rangle$ or $|m=-2\rangle$, both of their amplitudes will oscillate sinusoidally at the 4.040 GHz frequency. This corresponds to continuous tunneling back and forth between $|m=2\rangle$ and $|m=-2\rangle$ at twice the frequency, 8.080 GHz.
The odd-$m$ block, {$m= -1, +1$}, forms a separate pair of eigenstates (at $-15.06$ K and $-9.68$ K) that mix only with each other, and never with the even-$m$ block containing the qubit states. For this Hamiltonian, they play no role in the qubit's dynamics and remain fully decoupled.

\section{Hamiltonian Validation}
\label{sec:validation}

The Hamiltonian discussed in the previous section is a general zero-field-splitting model applicable to any ion described by axial and transverse crystal-field anisotropy, regardless of its specific spin, $D$, or $E$ values. We therefore validated our diagonalization methods by confirming that they correctly reproduced known independently-established energy spectra of other ions governed by the same Hamiltonian structure. We compared the Hamiltonian to a Fe$_8$ single-molecule magnet's previously established approximate benchmark and the Raman transition data for Fe(H$_2$O)$_6^{2+}$. We also implemented an independent Qiskit\cite{Qiskit}-based version of the proposed qubit system to compare the eigenvalues of the Hamiltonian.

\subsection{Qiskit Cross-Validation}
\label{subsec:qiskit}

To ensure that there were no implementation errors, we built the Hamiltonian independently using Qiskit\cite{Qiskit}, a quantum computing library, to compare the eigenvalues produced by both our analytic solution and Qiskit. To confirm whether there were any implementation errors, we rebuilt the Hamiltonian independently using Qiskit\cite{Qiskit}, a quantum computing library, to compare its eigenstates. Our 5-level ($S=2$) qubit system's Hamiltonian corresponds to a 3-qubit Pauli embedding in the Qiskit library. The two independent implementations produced identical eigenvalues: $-49.67$, $-49.48$, $-15.06$, $-9.68$, and 0.19 K, as well as the same energy span ($49.86$ K) and tunnel splitting ($\Delta = 0.19$ K). The analytic solution for these eigenvalues can be found in Appendix~\ref{app:S=2_analytic_solution}. This independent implementation doesn't validate our method choice, but rather shows that our method was implemented correctly.

\subsection{Fe$_8$ Single-Molecule Magnet}
\label{subsec:fe8}

Fe$_8$ is a single-molecule magnet with the formula [(tacn)$_6$Fe$_8$O$_2$(OH)$_{12}$]$^{8+}$, with a spin angular momentum of $S = 10$. We diagonalized the Hamiltonian (Eq~\ref{eq:hamiltonian_static}) to get the possible energy states of Fe$_8$, using the NumPy\cite{harris2020array} python library and its known parameters: $D = 0.292$\,K and $E = 0.046$\,K to calculate its energy barrier. Our eigenvalue calculations showed the energy barrier to be 32.6 K, while the experimentally accepted value is $\ \sim 29$\,K \ \cite{Jammalamadaka_2015}. The calculated value is 12\% higher than the literature-based value. This slight inaccuracy is expected since our model only includes second-order anisotropy ($D$ and $E$) while a real-world Hamiltonian includes fourth-order anisotropy terms. Nevertheless, this simplified system is in reasonable agreement with the previously established calculation for this compound's Hamiltonian. 

\subsection{Fe(H$_2$O)$_6^{2+}$}
\label{subsec:feH20}

The Fe(H$_2$O)$_6^{2+}$ system is a well characterized Fe$^{2+}$ system with experimentally measured electronic Raman transitions, which are inelastic light-scattering processes where a photon changes the electronic energy state of an atom or molecule rather than just its vibrational or rotational state. We applied the Hamiltonian diagonalization method to this compound using literature anisotropy values: $D = 17.19$\,K and $E = 0.947$\,K \cite{Krzystek2011}. The eigenvalue calculations predict transition values of 14.20, 19.88, 68.46, and 68.62 K, as shown in Appendix~\ref{subsec:validation_2_transition_energies}. The measured transition energies were 14.24,  20.00, and 68.49 K $\,$ \cite{Gnezdilov1991}. The first three predictions match up extremely well with the measured values. The last predicted value does not have a corresponding measured value; nevertheless, the three values match up very well, validating our diagonalization method to find the energy states of the system. Altogether, the three validations performed together demonstrate the method's physical validity and the correctness of the implementation. 

\section{Gate Control Schemes}
\label{sec:gate_control}

The Hamiltonian modeled in Sec.~\ref{sec:system_model} defines a five-level qubit encoding, but there isn't a mechanism to drive between the two lowest-energy states. In order for the system to represent a qubit, we must implement a driving mechanism to coherently move the population between $|\mathbf{0}\rangle$ (an approximate symmetric superposition of $|\pm 2\rangle$) and $|\mathbf{1}\rangle$ (the antisymmetric superposition of $|\pm 2\rangle$) and manipulate their relative phase. The driving mechanism must be able to move the in-phase $|\pm 2\rangle$ components into an out-of-phase state. These operators are conventionally known as the X-gate and the Z-gate. The X-gate is a qubit operator that fully transfers population from $|\mathbf{0}\rangle$ to $|\mathbf{1}\rangle$ or vice versa. The Z-gate is phase shift operator that leaves the state $|\mathbf{0}\rangle$ unchanged and maps the state $|\mathbf{1}\rangle$ to $-|\mathbf{1}\rangle$ \ \cite{NielsenChuang}. We explored numerous driving mechanisms before identifying a functional method.

\subsection{Switched magnetic field}
\label{subsec:switched_magnetic_field}

The initial design for the system involves switching a 5 T magnetic field on and off. At 5 T, the field's data leakage between the two qubit states is bounded by 0.01\% . The system undergoes natural tunnel splitting when the field is off and population stays static while the field is on. As shown in Appendix \ref{app:S=2_analytic_solution_zeeman}, the eigenvalues of the system with a 5 T magnetic (Zeeman) field term are $-62.99$, $-36.18$, $-19.60$, $-5.14$, and $0.21$ K. The energy gap between the two qubit basis states (now $|\mathbf{0}\rangle \approx |+2\rangle$ and $|\mathbf{1}\rangle \approx |-2\rangle$) is $ \Delta = 26.81$\, K. These two states are no longer degenerate, so the system does not undergo tunnel splitting. The angular frequency of the qubit transition, $\omega_0$, is given by 
\begin{equation}
\label{eq:omega_0}
\omega_0 = \Delta/\hbar \approx 3.51 \times 10^{12} \ \text{rad/sec}.
\end{equation}
The calculated X-Gate and Z-Gate transition times are 123.7 and 0.89 picoseconds, respectively. The Z-gate~\cite{NielsenChuang} is implemented as a free evolution of $t_Z = \pi /\omega_0 \approx 0.89$ ps in the 5 T field, which gives $|\mathbf{1}\rangle$ a relative phase of $\pi$ with respect to $|\mathbf{0}\rangle$. The caveat with this driving mechanism, however, is that it is currently impossible to switch a 5 T field on/off with picosecond precision, which is needed for the X and Z logic gates. 

\subsection{Symmetry-Forbidden mechanisms}
\label{subsec:symmetry_forbidden}

Two other explored methods were Single-tone (Sx) driving and Electric Dipole Spin Resonance (EDSR). Single-tone magnetic (Sx) driving oscillates an applied magnetic field along the crystallographic $x$-axis, the same transverse axis defined by the $E(\hat{S}_x^2 - \hat{S}_y^2)$ term in Eq.~\eqref{eq:hamiltonian_static}, synchronized with the qubit. The Single-tone (Sx) driving mechanism was modeled using
\begin{equation}
    \hat{H}(t) = \hat{H}_{\text{static}} + g_L\mu_B B_1 \cos(\omega_0t)\hat{S}_x,
\end{equation}
which contains coupling via the $\hat{S}_x$ operator, with $\omega_0 \approx 3.51\times10^{12}$ rad/s ($\hbar\omega_0 \approx 26.8$ K), resonant with the 5 T idle-field splitting.  The system is initialized in the $|\mathbf{0}\rangle$ state; $\hat{H}_{\text{static}}$ is the static-state Hamiltonian from Eq.~\eqref{eq:hamiltonian_static}; $B_1$ is the oscillating magnetic field amplitude. 

Electric Dipole Spin Resonance oscillates an electric field that indirectly drives the spin transitions through spin-orbit coupling; this is usually used in systems where magnetic driving is impractical. The EDSR driving mechanism was modeled using
\begin{equation}
\hat{H}(t) = \hat{H}_{\text{static}} + \Lambda \cos(\omega_0 t) \hat{S}_x ,
\end{equation}
where $\Lambda$ is an effective energy amplitude that stands for the electric-field coupling through spin-orbit interaction. The rest of the parameters stay the same as described for the Single-tone driving mechanism.

Since $\hat{S}_x$ is a $\Delta m = 1$ operator by construction, the issue with both methods is that they can only bridge a $\Delta m = 1$ gap, while the proposed qubit system requires coherent driving across $\Delta m = 4$. Single-tone $\hat{S}_x$-driving showed a $\sim$0.0001 population transfer at a drive amplitude of $B_1 = 0.01$ T and didn't increase even as the magnetic field strength increased to $B_1$ = 1.0 T, demonstrating that the limitation was not due to insufficient field strength. Further, the matrix element $\langle \mathbf{0}| \hat{S}_x |\mathbf{1}\rangle = 0$, indicating that the $\Delta m = 1$ selection rule is the limiting factor for both driving mechanisms.

\subsection{Two-photon Driving}
\label{subsec:two_photon}

The two photon driving mechanism uses the same $\hat{S}_x$-coupled drive Hamiltonian introduced above, while oscillating at half the resonant frequency, $\omega_{\text{2-photon}}$ = $\omega_0$/2, requiring the two photons to be absorbed at once. This is a second-order process, with a non-zero effective matrix element for $\Delta m$ = 4, as opposed to the first-order single-photon process case, where the direct matrix element for $\Delta$m = 4 is zero. A two-photon process allows the intermediary state to mediate coupling between the two lowest-energy states, circumventing the single-photon selection rule. We swept the drive amplitude $B_1$ from 0.01 T to 5.0 T, initialized in the $|\mathbf{0}\rangle$ state. The simulation showed a maximum population transfer of only 7.3\%, peaking at $B_1$ $\approx$ 0.5 T. Increasing $B_1$ beyond this point reduced rather than improved the transfer. One possible reason for this decline is the AC Stark shift, which shifts the system's energy levels as a function of drive strength. This is a problem because the drive frequency, $\omega_{\text{2-photon}}$, was fixed using the static (zero-field) energy gap, so once the AC Stark shift moves the levels, the drive falls out of resonance at higher field strength. 

\subsection{Crystal-strain Driving}
\label{subsec:crystal-strain}

In this scheme, the 5 T magnetic field remains on continuously. The crystal-strain driving mechanism oscillates the crystal's strain, stretching and compressing the lattice around the ion through a surface acoustic wave (SAW). This strain is assumed to modulate the transverse anisotropy by $\pm 10$ \%, 
\begin{equation}
E(t) = E[(1 + 0.1\cos(\omega_0 t)] ,
\end{equation}
where $\omega_0 \approx 3.51 \times 10^{12}$ rad/s is resonant with the ~26.8 K ($\approx558.4$ GHz) gap between the lowest-energy states. The oscillation supplies quanta of energy $\hbar\omega_0$ equal to the gap, so it drives transitions between $|\mathbf{0}\rangle$ and $|\mathbf{1}\rangle$. 

The $E$-term only connects states differing by $\Delta m = \pm 2$. It therefore has no direct matrix element between $|m=+2\rangle$ and $|m=-2\rangle$ (which differ by $\Delta$m = 4), and couples them only virtually, through m = 0. However, because of the $E$-term, at 5 T, each $m \approx \pm 2$ eigenstate contains a small admixture of  $m = 0$, and the 5 T field further mixes the symmetric and antisymmetric combinations. As a result, the matrix element of $\hat{S}_x^2 - \hat{S}_y^2$ between the two qubit eigenstates at 5 T is given by $\langle \mathbf{0}|  \hat{S}_x^2 - \hat{S}_y^2 |\mathbf{1}\rangle \approx \langle {-2}|  \hat{S}_x^2 - \hat{S}_y^2 |{2}\rangle \approx 0.23$. This gives an effective coupling of about $0.01$ K and a $\pi$-pulse time of $ \approx 1150$ ps. The strain therefore flips the spin through this small $m=0$ component, and the steady field holds the new state once the drive is switched off. 

This method produces a complete transfer of population between the two lowest-energy states (X-Gate) and a $\pi$ phase accumulation (Z-Gate). The X-gate takes 1148 ps with a fidelity of $99.5\%$, and the Z-gate takes ~1149 ps with a fidelity of $99.7\%$. Here fidelity measures how closely the final quantum state matches the intended target state, with a value of 1 indicating a perfectly executed gate. The meaning of these values and their derivations are discussed further in the following section.

\section{Population shift and phase accumulation}
\label{sec:population_shift_and_phase_accumulation}

A quantum computer needs to be able to efficiently shift population from one state to another and accumulate phase in the direction perpendicular to the population shift. As previously established, the best method to drive the population is by using the crystal-strain driving mechanism described in subsection~\ref{subsec:crystal-strain}. For phase accumulation, we use a virtual phase tracker discussed in more detail in subsection~\ref{subsec:phase accumulation}.

\subsection{Population Shift}
\label{subsec:population_shift}

\begin{figure}[htbp]
    \centering
    \includegraphics[width=0.8\linewidth]{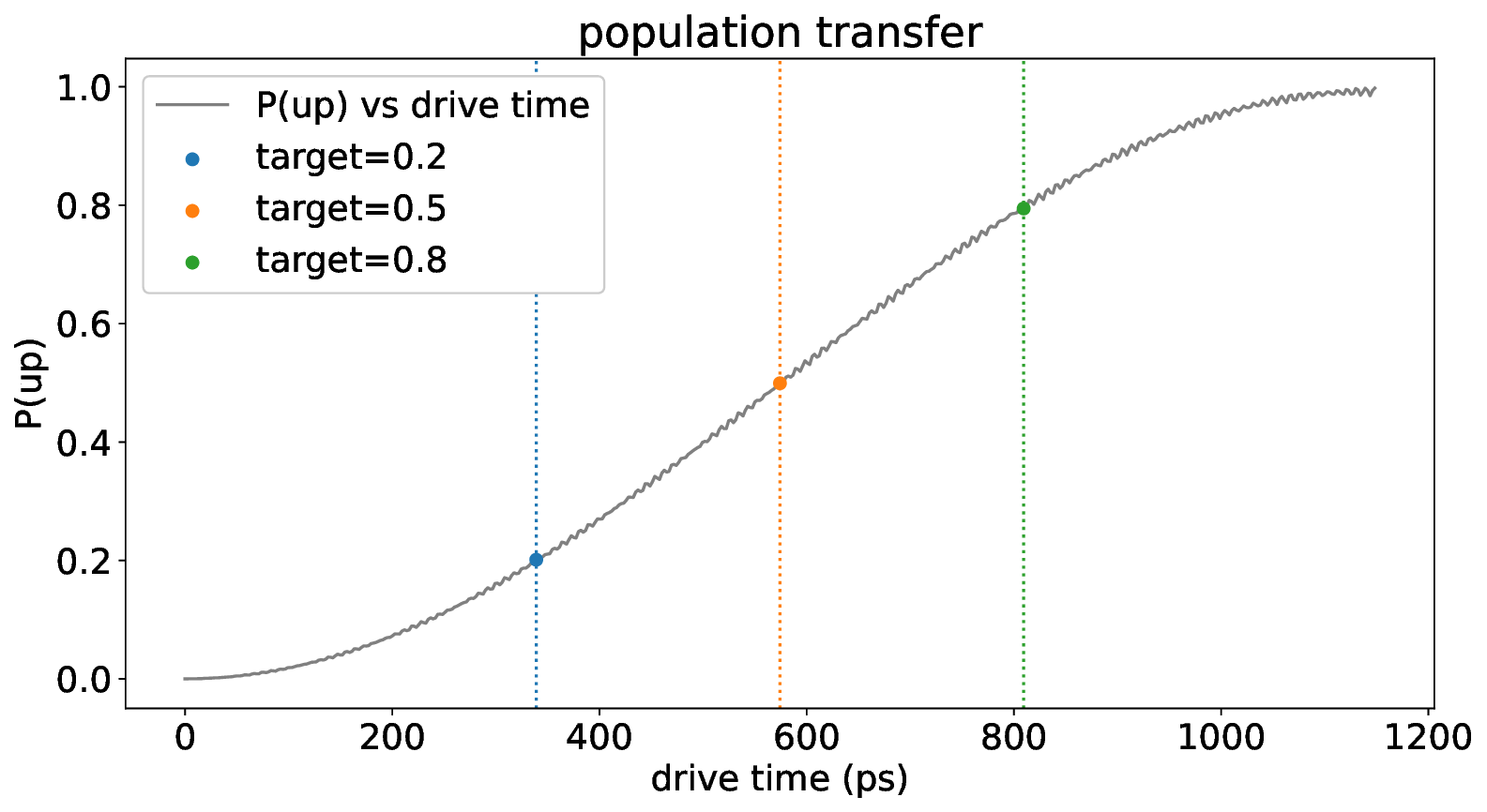}
    \caption{The population shift as an oscillating strain is applied to the crystal lattice surrounding the ion over time. P(up) represents the $|\mathbf{1}\rangle$ state} \label{fig:population_evolution_with_strain_driving}
\end{figure}

To switch between the two lowest-energy states, $|\mathbf{0}\rangle$ to $|\mathbf{1}\rangle$ or vice versa, we perform an X-Gate (the qubit bit flip)~\cite{NielsenChuang}. In our system this corresponds to a $\pi$ pulse of the resonant drive, which fully transfers the population between the two states. We used the crystal-strain driving mechanism discussed in subsection~\ref{subsec:crystal-strain}, as well as a continuous $5\text{ T}$ magnetic field, discussed in subsection~\ref{subsec:switched_magnetic_field}, to hold the qubit's state to prevent natural tunnel-driven oscillations. To characterize the system's evolution, we solved the time-dependent Schrödinger equation.  
\begin{equation}
i\hbar \frac{\partial}{\partial t}|\psi(t)\rangle = \hat{H}(t)|\psi(t)\rangle
\end{equation}
with our specific time-dependent Hamiltonian: 
\begin{equation}
\hat{H}(t) = \hat{H}_{\text{static}} + A \cos(\omega_0 t)(\hat{S}_x^2 - \hat{S}_y^2)
\end{equation}
with $\omega_0 \approx 3.51 \times 10^{12}$  rad/s (with $\hbar\omega_0 \approx 26.8$ K), resonant with the idle-field splitting; drive amplitude $A = 0.1E \approx 0.09$ K, and $\hat{H}_{\text{static}}$ being the static Hamiltonian from Eq.~\eqref{eq:hamiltonian_static}. The system is initialized in the $|\mathbf{0}\rangle$ state and we track population transfer into $|\mathbf{1}\rangle$ over time. We compute the population of $|\mathbf{0}\rangle$ and $|\mathbf{1}\rangle$ by solving the time-dependent Schrödinger equation. The time evolution of these population probabilities is plotted in Figure~\ref{fig:population_evolution_with_strain_driving}.

The X-Gate duration, defined as the drive time required for the strain-drive-induced Rabi oscillation to fully transfer population from $|\mathbf{0}\rangle$ to $|\mathbf{1}\rangle$ ($p_{target}$ = 1), was simulated to be approximately $1148$\text{ ps} and some of the intermediary states are listed in Table~\ref{tab:population_shift_times}.
\begin{table}[htbp]
    \centering
    \caption{The times, in picoseconds, to reach specific probabilities of $|0\rangle$ occupation, comparing direct simulation results against predictions from half the theoretical Rabi period, Eq.\ (\ref{Rabi}).}
    \label{tab:population_shift_times}
    \begin{tabular}{|c|c|c|c|}
        \hline
        \textbf{Probability $|\mathbf{0}\rangle$} & \textbf{simulated (ps)} & \textbf{Predicted (ps)} & \textbf{\% error} \\ \hline
        0.2 & 339 & 339 & 0.0 \\ \hline
        0.5 & 574 & 575 & 0.002 \\ \hline
        0.8 & 809 & 810 & 0.001 \\ \hline
        1.0 & 1148 & 1149 & 0.0009 \\ \hline
    \end{tabular}
\end{table}
These simulated durations match up almost exactly, with only a $0.091\%$ error, with half of the theoretical Rabi Period ($T/2$) (labeled Predicted in Table \ref{tab:population_shift_times}): 
\begin{equation}
\label{Rabi}
T = \frac{2\pi}{\Omega_R}
\end{equation}
where:
\begin{equation}
\Omega_{R} = \frac{A |M|}{\hbar}
\end{equation}
with $A$ being the strain amplitude and $M = \langle \mathbf{0}|\hat{S}_{x}^{2} - \hat{S}_{y}^{2}|\mathbf{1}\rangle$. Consequently, the required drive duration to reach a specific population shift can be accurately predicted to determine the application time of the strain drive:
\begin{equation}
t = \frac{2}{\Omega_R} \arcsin\left(\sqrt{p_{\text{target}}}\right),
\end{equation}
with $p_{target}$ being the probability of the $|\mathbf{1}\rangle$ state when starting from the $|\mathbf{0}\rangle$ state.

This process allows the qubit system to smoothly shift population through a strain driving mechanism with a constant magnetic field. Continuing the drive beyond the X-Gate duration returns the population to $|\mathbf{0}\rangle$ after a second half-period, consistent with sinusoidal Rabi oscillation. Assuming the spin-phonon coupling constant $G$ (the strength of the ion's coupling to lattice vibrations, discussed in detail in Sec.~\ref{sec:temperature_and_decoherence}) is approximately equal to $144$\, K at a temperature of 10 K, the X-Gate had a fidelity of 0.995. These temperature and $G$ values are discussed further in Sec.~\ref{sec:temperature_and_decoherence}. The potential methods of readout of these results from the population are discussed further in Sec.~\ref{sec:readout}.

\subsection{Phase Accumulation}
\label{subsec:phase accumulation}

Phase accumulation is often used in qubit systems to shift the qubit's state to enable logic gate operations without changing the overall energy. For example, phase can be shifted in a way that cancels out wrong answers and boosts the correct answer through interference. A Z-Gate is a phase-flip operator that leaves the state $|\mathbf{0}\rangle$ unchanged and maps the state $|\mathbf{1}\rangle$ to $-|\mathbf{1}\rangle$ \ \cite{NielsenChuang}. It is executed by applying a half-pulse (starting from either state), creating an equal superposition followed by a hold period, where the phase accumulates, and then another half-pulse; this method is known as the Ramsey interferometry approach. In a qubit system, phase is known to accumulate from natural energy splitting, most when in an equal superposition, with no phase accumulation at either singular state. 
The total gate time for the Z-Gate was simulated to be 1149$\text{ ps}$, with a fidelity of 0.997, assuming $G$ $\approx$ $D$ at 10 K. These parameters are discussed further in Sec.~\ref{sec:temperature_and_decoherence}.

Phase accumulation during the hold period is problematic because when performing multiple back-to-back operations, such as the X-gate and the Z-gate, the pulses need to know the correct phase to start from; but, the correct phase reference can shift depending on the hold duration. In order to combat this issue, we implemented a virtual phase tracker that keeps track of the phase of the qubit system externally in a classical computer and not on the physical system. This is a very common practice among industry-level quantum computers, with many reporting the use of virtual phase tracking.

\section{Decoherence and Temperature Robustness}
\label{sec:temperature_and_decoherence}

When the Fe$^{2+}$ ion sits in a crystal lattice, it is subject to thermal noise or vibrations, whose quanta are called phonons. Phonons are quantum packets of vibrational energy that move through a solid material such as crystals. Phonons modulate the crystal field at the ion through the lattice strain $\epsilon$. We denote the coefficient coupling the transverse anisotropy $E$ to the strain as $G_E=\partial E/\partial\epsilon$; the same coefficient sets the strain-drive amplitude, $A = G_E\,\epsilon_{\mathrm{ac}}$, where $\epsilon_{\mathrm{ac}}$ is the amplitude of the applied ac strain and $A$ is the drive amplitude in the modulation $E(t)$. Strain also shifts the axial term $D$ through $\partial D/\partial\epsilon$, but $\hat S_z^2$ is nearly diagonal in the qubit basis, matrix element $\langle \mathbf{0}|  \hat{S}_z^2 |\mathbf{1}\rangle\approx 0.008$ versus $\langle \mathbf{0}|  \hat{S}_x^2 - \hat{S}_y^2 |\mathbf{1}\rangle\approx 0.23$, so $D$-type strain couples to the qubit transition roughly thirty times more weakly than $E$-type strain per unit coefficient. We therefore take the $E$-type coefficient $G_E$ as the relevant coupling. The two coefficients involve different strain components and are not constrained to be equal. To the best of our knowledge, no value has been reported for $G_E$ or $\partial D/\partial\epsilon$ in Fe$^{2+}$:forsterite. We sweep values of $G_E$, in kelvin per unit strain, to find the range over which the qubit relaxes slowly compared to the gate execution time. For a non-Kramers ion the direct-process relaxation rate is\cite{AbragamBleaney1970}:
\begin{equation}
\gamma_0 = \frac{3\,(G_E\,M)^2\,\Delta^3}
{2\pi \rho \hbar^4 v_s^5}\,\coth\!\left(\frac{\Delta}{2T}\right),
\end{equation}
where $M =|\langle\mathbf{0}|\hat S_x^2-\hat S_y^2|\mathbf{1}\rangle| \approx 0.23$ is the matrix element of the strain operator between the qubit states, obtained by diagonalizing the Hamiltonian at 5 T. In the equation, $\Delta = 26.81$\,K, $\rho = 3271$ kg/m$^3$, $T = 10$\,K and $v_s \approx 5.56$ km/s is the Debye-averaged sound velocity. We take $1/\gamma_0$ as the relaxation time. The criterion for the sweep is to check whether the system's relaxation time is greater than 100 X-gate execution times. This holds for $G_E \lesssim 255$\,K per unit strain, considering only the direct-process relaxation, shown in Table~\ref{tab:G_sweep_10K}. Cr$^{2+}$:CdTe, another $S = 2$ spin-orbit system, has a shear coupling corresponding to roughly $114\pm46$\,K \cite{besombes2019optical} per unit strain, which lies within the window. While this does not tell us the value for $G_E$, it suggests that the window is not unreasonable for an $S = 2$ system. Nevertheless, $G_E$ remains a quantity that needs to be measured.

\begin{table}[htbp]
\centering
\caption{Gate margin as a function of the spin-phonon coupling constant $G_E$, swept across seven orders of magnitude, evaluated at $T=10$~K. Margin is defined as the ratio of the direct-process relaxation time to the gate operation time. Only representative points are shown; the design remains SAFE ($>100\times$ margin) up to $G_E = 255.00$ and does not FAIL ($<1\times$ margin) until $G_E = 2511.89$.}
\label{tab:G_sweep_10K}
\begin{tabular}{cccc}
\toprule
$G_E \,(K)$ & $\gamma_0$ (Hz) & $X$-gate margin & Status \\

$1.00$   & $1.34\times10^{2}$ & $6.5\times10^{5}\times$ & SAFE \\
$50.12$    & $3.36\times10^{5}$  & $2591\times$ & SAFE \\
$125.9$   & $2.12\times10^{6}$  & $410.4\times$              & SAFE \\
$255.0$   & --                  & $100\times$              & \textit{(safe boundary)} \\
$501.2$   & $3.36\times10^{7}$  & $25.9\times$             & MARGINAL \\
$1000.$   & $1.38\times10^{7}$  & $6.5\times$              & MARGINAL \\
$2512$   & --                  & $1.0\times$              & \textit{(fail boundary)} \\
$3162$   & $1.34\times10^{9}$  & $0.7\times$              & FAILS \\

\end{tabular}
\end{table}

When assuming $G_E \approx 144$\,K, the decoherence rate is calculated to be $2.77$ x $10^6$ Hz. We plotted the infidelity of the logic gates with decoherence rate at 10 K in Figure 2.
\begin{figure}[htbp]
    \centering
    \includegraphics[width=0.8\linewidth]{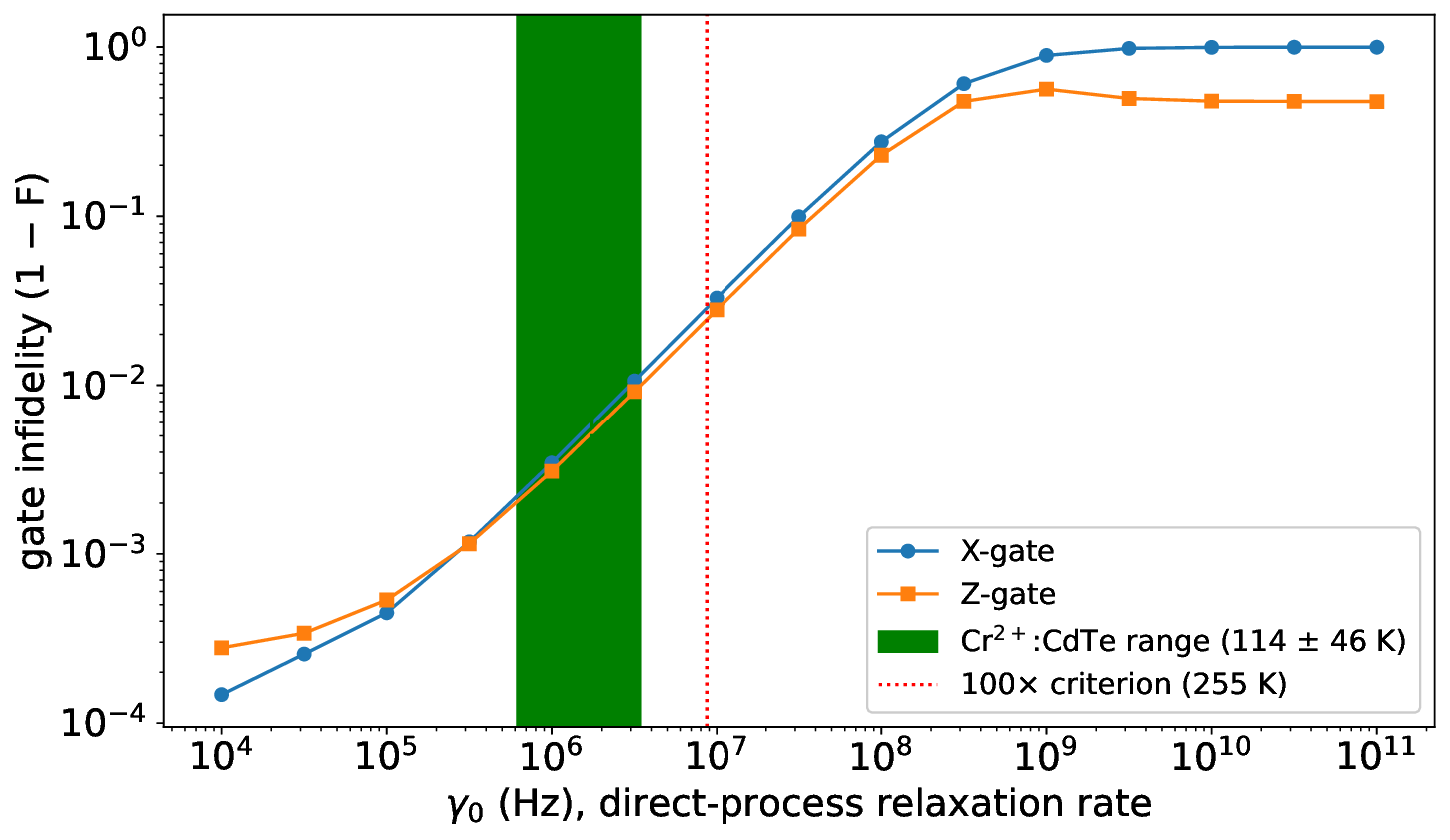}
    \caption{The strain-drive gate infidelity of the system vs decoherence rate at T = 10 K. The figure shows that at estimated decoherence rate the infidelity of the X and Z Gate logic gates are extremely low. } 
    \label{fig:graph_2}
\end{figure}
We also plotted the number of X-Gate margins executable before complete decoherence with decoherence rate at 10 K in Figure \ref{fig:graph_3}, using 
\begin{equation}
\text{margin} = \frac{1}{\gamma_0 \cdot t_{\text{gate}}}.
\end{equation}
\begin{figure}[htbp]
    \centering
    \includegraphics[width=0.8\linewidth]{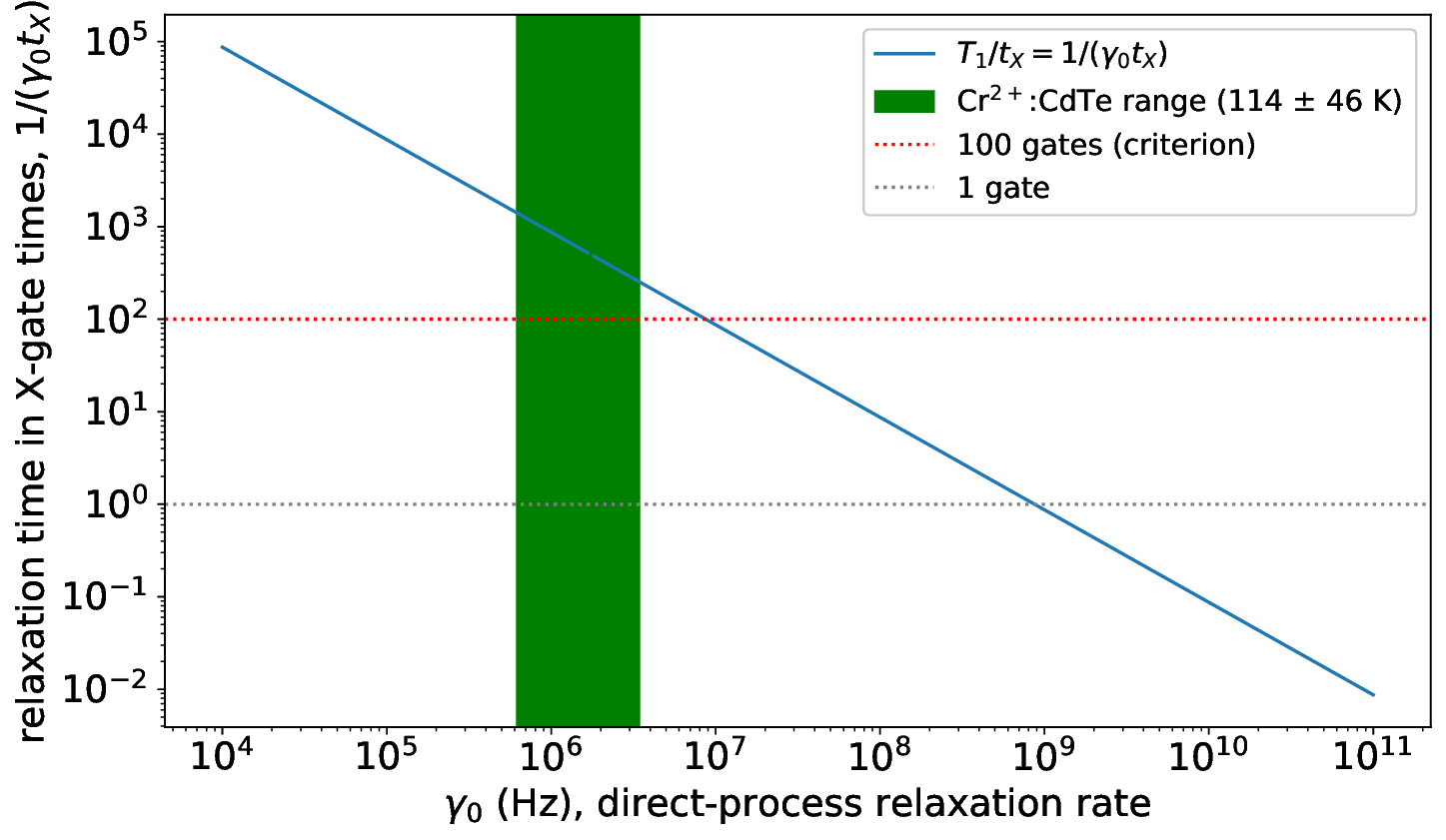}
    \caption{The X-Gate budget vs. decoherence rate at T = 10 K. The figure shows that for direct-process relaxation rates $ \ \approx 10^7$ the system can safely execute 100+ X Gates}
    \label{fig:graph_3}
\end{figure}
We estimated the effect of temperature on the system and found that at temperatures $<$ 10 K,  population leakage is shown to be $<$ 1\%. These results are shown in more detail in Table \ref{tab:thermal_leakage}, calculated using the Boltzmann distribution ratio: 
\begin{equation}
\frac{P(m \approx 0)}{P(\text{ground doublet})} \approx \exp\left(-\frac{\Delta_{\text{span}}}{T}\right).
\end{equation}
\begin{table}[htbp]
\centering
\caption{Thermal population leakage into the intermediate $m=0$ crystal-field level as a function of temperature, estimated via a Boltzmann factor $\exp(-\Delta_{\text{span}}/T)$ using the energy span of 49.86~K.}
\label{tab:thermal_leakage}
\begin{tabular}{cc}
\toprule
Temperature (K) & Thermal leakage into $m=0$ \\
1  & 0.00\% \\
4  & 0.00\% \\
5  & 0.00\% \\
7  & 0.08\% \\
8  & 0.20\% \\
10 & 0.68\% \\
15 & 3.60\% \\
20 & 8.26\% \\
25 & 13.60\% \\
\end{tabular}
\end{table}
There are two caveats to these values as they only compare two levels $m = \pm 2$ and $m = 0$. A more accurate account would include all five energy levels. Additionally, the formula used only considers the direct-process relaxation, but it doesn't consider two other possible relaxation channels, Raman and Orbach. They can be calculated using 
\begin{equation}
\Gamma_{\text{R}}(T) = C_{\text{R}} T^n, \qquad \Gamma_{\text{O}}(T) = \frac{A_{\text{O}}}{\exp(\Delta /  T) - 1}
\end{equation}
Where $\Gamma_{\text{R}}$ is the Raman relaxation, $\Gamma_{\text{O}}$ is the Orbach relaxation, and $n$ is the temperature exponent characterizing the power-law dependence of Raman spin-lattice relaxation (conventionally considered to be $n \approx 7$ for $S=2$ non-Kramers ions in low-temperature acoustic-phonon models). Unfortunately, the two prefactor terms needed: $C_{\text{R}}$ and $A_{\text{O}}$, for Fe$^{2+}$:forsterite remain unmeasured. The calculations for the Raman and Orbach relaxations might lower the duration for which the qubit remains functional; this remains a limitation in our work and awaits further calculations to measure the absolute time-duration for which the qubit would be functional.

The margin table and gate simulations show that decoherence is not a major issue at temperatures $<$ 10 K, and Table \ref{tab:thermal_leakage}, thermal leakage, shows that the qubit's two-level structure itself stays intact up to 10 K. The direct-process decoherence measurements show that all temperatures less than 100 K have a large enough margin to execute 1,000 X-Gates; however, we are limiting our claim to 10 K due to the leakage of information exceeding one percent beyond that point. Non-Kramer's ions are significantly more susceptible to thermal noise than Kramer's ions because they are not subject to Kramer's theorem \cite{Kramers1930}, which states that ions with an odd number of electrons have doubly degenerate states. Nevertheless, the decoherence test proves that despite this disadvantage, the proposed qubit system remains degenerate enough to act as a functional qubit system.

\section{Readout}
\label{sec:readout}

Everything discussed above has been simulated in the qubit's internal dynamics. However, for this information to be useful, we must be able to convert these states and superpositions into classical real-world values. As this paper is purely theoretical, we are not able to test readout methods experimentally. However, we propose two potential readout methods that could work for this qubit system: Inductive/magnetometric readout and Magnetic resonance force microscopy (MRFM) readout.

\subsection{Inductive/magnetometric readout}
\label{subsec:Inductive_readout}

Inductive magnetometric detection is based on coupling the magnetic moment of the ions with either a superconducting quantum interference device or a resonant LC circuit to measure the flux difference of the two lowest-energy states. This process may be achieved by bringing the spins into geometric proximity of the superconducting pick-up loop or micro-fabricated inductor. Here, the spins behave like small bar magnets and emanate magnetic field lines. Some of these magnetic field lines enter the open portion of the inductor coil. The magnetic flux linked to the coil is the surface integral of the field:
\begin{equation}
\Phi = \iint \vec{B}_{\rm spin} \cdot d\vec{A}.
\end{equation}
In the presence of the continuous 5 T field used in our scheme, the states $|\mathbf{0}\rangle$ and $|\mathbf{1}\rangle$ are localized, with $\langle S_z\rangle \approx -2$ and $+2$, respectively. They therefore have opposite magnetic dipole orientations, so the state of the qubit changes the total magnetic flux measured by the coil. 

\subsection{Magnetic resonance force microscopy (MRFM)}
\label{subsec:MRFM_readout}

Magnetic resonance force microscopy uses a mechanical cantilever to measure the minute force exerted by the spin of the ion, which helps in the detection of the force created between the magnetic tip and the spin state that is in the sub-attonewton range. 
The magnetic tip is mounted on a vibrating ultra-soft micro-cantilever, and its coupling to the ion depends on $\langle S_z\rangle$. Because the cantilever motion (kHz-MHz) is many orders of magnitude slower than the 26.8 K ($\approx 559$ GHz) gap between the qubit states, and the interaction is diagonal in the $|\mathbf{0}\rangle$, $|\mathbf{1}\rangle$ basis, the readout does not drive transitions and the qubit state is preserved. The resulting spin-dependent force on the cantilever changes its resonance frequency by an amount that depends on the state of the qubit. This frequency shift is continuously monitored by an optical interferometer. 

These are just two proposals for readout methods, which are not guaranteed to work by any means. However, they carry a huge caveat, as the proposed readout methods require cooling to a lower temperature than what was established in Sec.~\ref{sec:temperature_and_decoherence} in order to operate, slightly undermining their advantages.

\section{Conclusion}
\label{sec:conclusion}

We considered a single Fe$^{2+}$:forsterite qubit system that naturally undergoes tunnel splitting across $m = \pm 2$, at 4.040 GHz. We implemented three independent Hamiltonian validation methods to validate our process in creating the Hamiltonian. Our model includes a continuous 5 T magnetic field, which prevents the system from unintentionally shifting the population through tunnel splitting. In order to drive transitions between $|\mathbf{0}\rangle$ and $|\mathbf{1}\rangle$, we simulated a crystal-strain driving mechanism, which oscillates the crystal's strain, stretching and compressing the lattice around the ion through a surface acoustic wave (SAW); this oscillation supplies quanta of energy $\hbar\omega$, equal to the gap, $\hbar \omega_0$. We use a virtual phase tracker to keep track of phase accumulation within the system since phase unintentionally accumulates under a 5 T magnetic field. Both the X and Z logic gates run at 1.15 ns with fidelity $>$ 99.7\%, and the decoherence margin remains large enough ( $>$ 10,000 X-Gate margins) up to at least 10 K, meaning no dilution fridge is required, unlike many qubit systems. Limitations to this model include an unmeasured spin-phonon coupling constant ($G_E$) value, Raman/Orbach relaxation not considered as their prefactor for Fe$^{2+}$:forsterite remain unmeasured, and lack of an experimentally tested readout method. These open up opportunities for future work measuring the true spin coupling constant ($G_E$) for this system, experimentally measuring the Raman/Orbach prefactor terms to then calculate their relaxation effects on the system, and experimentally testing the proposed and other readout methods. The diagonalization method used to find the energy states of the system can also be applied to ions with different spin angular moments, such as $S = 1,\ 3/2,\ 2$, and possibly more, 
to determine the energy states of other similar ions and evaluate their functionality as effective qubits.   

\section{Author Declarations}
\subsection{Conflict of Interest}
The authors have no conflicts to disclose.

\subsection{Author Contributions}
\textbf{Dishaanth S. Pillai:} Conceptualization (lead); Formal analysis (supporting); Investigation (lead); Software (lead); Writing - original draft (lead); Writing - review \& editing (equal). \textbf{G. M. Wysin:} Conceptualization (supporting); Formal analysis (lead); Investigation (supporting);  Writing - original draft (supporting);  Writing - review \& editing (equal).

\section{Data Availability}
The data that support the findings of this study are available from Dishaanth S. Pillai upon reasonable request.





\appendix

\section{$\mathbf{S=2}$ Diagonalization}
\label{app:S=2_analytic_solution}
The spin Hamiltonian applies to an ion of interest with net electron spin $S=2$ such as $\mathrm{Fe}^{2+}$ embedded in forsterite. It includes two forms of anisotropy with energy parameters $D$ and $E$ and could include an applied magnetic field $B$ along the $z$-axis. We believe it applies to the ion of interest sitting in the crystal field of neighboring ligand ions along the $z$-direction, and along the $x$ and $y$ directions, which interact via dipole-dipole interactions with the ion of interest. The zero-field-splitting Hamiltonian for a quantum spin is
\begin{equation}
\hat{H}=D \hat{S}_{z}^{2}+E\left(\hat{S}_{x}^{2}-\hat{S}_{y}^{2}\right)
\end{equation}
$\hat{H}$ can be represented in eigenstates of $\hat{S}_{z}$, which are denoted $|m\rangle$, where $m=2,1,0,-1,-2$ are the eigenvalues of $\hat{S}_{z}$ for $S=2$. The $D$-interaction depends on $m$ directly; it is diagonal in states of $\hat{S}_{z}$. The $E$-interaction term is not expressable using $m$; it changes the value of $m$. This can be seen by writing it using raising and lowering operators, defined by direct and inverse relations,
\begin{alignat}{2}
\hat{S}^{+} &= \hat{S}_{x} + i\hat{S}_{y}, &\qquad \hat{S}_{x} &= \frac{1}{2}\left(\hat{S}^{+} + \hat{S}^{-}\right), \nonumber \\
\hat{S}^{-} &= \hat{S}_{x} - i\hat{S}_{y}, &\qquad \hat{S}_{y} &= \frac{1}{2i}\left(\hat{S}^{+} - \hat{S}^{-}\right).
\end{alignat}
Some algebra gives
\begin{alignat}{2}
& \left(\hat{S}^{+}\right)^{2}=\hat{S}_{x}^{2}-\hat{S}_{y}^{2}+\mathrm{i}\left(\hat{S}_{x} \hat{S}_{y}+\hat{S}_{y} \hat{S}_{x}\right), \nonumber \\
& \left(\hat{S}^{-}\right)^{2}=\hat{S}_{x}^{2}-\hat{S}_{y}^{2}-\mathrm{i}\left(\hat{S}_{x} \hat{S}_{y}+\hat{S}_{y} \hat{S}_{x}\right), \quad \Longrightarrow \quad\left(\hat{S}^{+}\right)^{2}+\left(\hat{S}^{-}\right)^{2}=2\left(\hat{S}_{x}^{2}-\hat{S}_{y}^{2}\right). 
\end{alignat} 
Then the zero-field Hamiltonian is
\begin{equation}
\label{QHam}
\hat{H}=D \hat{S}_{z}^{2}+\frac{E}{2}\left[\left(\hat{S}^{+}\right)^{2}+\left(\hat{S}^{-}\right)^{2}\right]
\end{equation}
This does not conserve $m$, but rather, the $E$-term changes $m$ by $\pm 2$. This is because the raising and lowering operators have the properties:
\begin{alignat}{4}
\hat{S}^{+}|m\rangle &= \sqrt{S(S+1)-m(m+1)}|m+1\rangle, &
(\hat{S}^{+})^{2}|m\rangle &=\ \sqrt{\cdots}\sqrt{\cdots}|m+2\rangle, \nonumber\\
\hat{S}^{-}|m\rangle &= \sqrt{S(S+1)-m(m-1)}|m-1\rangle, &
(\hat{S}^{-})^{2}|m\rangle &= \sqrt{\cdots}\sqrt{\cdots}|m-2\rangle
\end{alignat}
where $\sqrt{\cdots}$ indicates the appropriate raising or lowering factors.
Consider a state with $m= \pm 1$. When $\hat{H}$ acts on it, it produces components with $m=1$ and $m=-1$, and nothing else, because there are no states with $m= \pm 3$ for $S=2$. Instead, consider a state with $m=0$. When $\hat{H}$ acts on it, the result will be a combination of $m=0$ with $m= \pm 2$ and nothing else. This shows that the $m= \pm 1$ or odd subspace is separated from the $m=0, \pm 2$ or even subspace. Therefore, $\hat{H}$ can be diagonalized by solving a 2 × 2 matrix for the odd subspace and a 3 × 3 matrix for the even subspace.

For the $m= \pm 1$ subspace problem, we have
\begin{align}
\left(\hat{S}^{+}\right)^{2}|1\rangle=0, &&\left(\hat{S}^{-}\right)^{2}|1\rangle=\hat{S}^{-} \sqrt{6}|0\rangle=\sqrt{6} \sqrt{6}|-1\rangle=6|-1\rangle, \nonumber \\
\left(\hat{S}^{-}\right)^{2}|-1\rangle=0, & &\left(\hat{S}^{+}\right)^{2}|-1\rangle=\hat{S}^{+} \sqrt{6}|0\rangle=\sqrt{6} \sqrt{6}|1\rangle=6|1\rangle . \hskip 0.43cm
\end{align}
An eigenstate of $\hat{H}$ is written as a superposition of kets, or a column vector,
\begin{equation}
|\psi\rangle=a_{1}|1\rangle+a_{-1}|-1\rangle, \quad \text { or } \quad \psi=\binom{a_{1}}{a_{-1}} . 
\end{equation}
The constants $a_{1}$ and $a_{-1}$ are the (complex) amplitudes of the $m= \pm 1$ components. The eigenvalue problem in this subspace is written as $\hat{H}|\psi\rangle=\lambda|\psi\rangle$, or in matrix form, with $\lambda$ being an energy eigenvalue,
\begin{equation}
(\hat{H})(\psi)=\left(\begin{array}{cc}
D & 3 E  \\
3 E & D
\end{array}\right)\binom{a_{1}}{a_{-1}}=\lambda\binom{a_{1}}{a_{-1}}=\lambda(\psi) .
\end{equation}
The solutions are found from the determinant of $\hat{H}-\lambda \hat{I}$ being zero,
\begin{equation}
\left|\begin{array}{cc}
D-\lambda & 3 E  \\
3 E & D-\lambda
\end{array}\right|=(D-\lambda)^{2}-9 E^{2}=0 \quad \Longrightarrow \lambda=\lambda_{1}^{ \pm}=D \pm 3 E .
\end{equation}
For $\lambda_{1}^{+}=D+3 E$, the eigenvector is found from $(\hat{H}-\lambda \hat{I})(\psi)=0$,
\begin{equation}
\left(\begin{array}{cc}
D-(D+3 E) & 3 E  \\
3 E & D-(D+3 E)
\end{array}\right)\binom{a_{1}}{a_{-1}}=\left(\begin{array}{cc}
-3 E & 3 E \\
3 E & -3 E
\end{array}\right)\binom{a_{1}}{a_{-1}}=\binom{0}{0},
\end{equation}
which shows that $a_{1}=a_{-1}$ and an unnormalized eigenvector in terms of the $|m\rangle$ states is
\begin{equation}
\left|\psi_{1}^{+}\right\rangle=|1\rangle+|-1\rangle \quad \text { (symmetric). }
\end{equation}
For $\lambda_{1}^{-}=D-3 E$, the eigenvector is found from
\begin{equation}
\left(\begin{array}{cc}
D-(D-3 E) & 3 E \\
3 E & D-(D-3 E)
\end{array}\right)\binom{a_{1}}{a_{-1}}=\left(\begin{array}{cc}
3 E & 3 E \\
3 E & 3 E
\end{array}\right)\binom{a_{1}}{a_{-1}}=\binom{0}{0},
\end{equation}
which shows that $a_{1}=-a_{-1}$ and an unnormalized eigenvector is
\begin{equation}
\left|\psi_{1}^{-}\right\rangle=|1\rangle-|-1\rangle \quad \text { (antisymmetric). } 
\end{equation}
Of course, these states are not eigenstates of $\hat{S}_{z}$, so they can't be referred to using $m$. But interesting relations exist and they \emph{are} eigenstates of $\hat{S}_{z}^{2}$,
\begin{equation}
\hat{S}_{z}\left|\psi_{1}^{+}\right\rangle=\left|\psi_{1}^{-}\right\rangle, \quad \hat{S}_{z}\left|\psi_{1}^{-}\right\rangle=\left|\psi_{1}^{+}\right\rangle, \quad \hat{S}_{z}^{2}\left|\psi_{1}^{+}\right\rangle=\left|\psi_{1}^{+}\right\rangle, \quad \hat{S}_{z}^{2}\left|\psi_{1}^{-}\right\rangle=\left|\psi_{1}^{-}\right\rangle . 
\end{equation}
This implies that they both have zero expectation values for $\hat{S}_{z}$, because the states are orthogonal,
\begin{equation}
\left\langle\psi_{1}^{+}\right| \hat{S}_{z}\left|\psi_{1}^{+}\right\rangle=0, \quad\left\langle\psi_{1}^{-}\right| \hat{S}_{z}\left|\psi_{1}^{-}\right\rangle=0 . 
\end{equation}
The expectation values for $\hat{S}_{z}^{2}$ are nonzero:
\begin{equation}
\frac{\left\langle\psi_{1}^{+}\right| \hat{S}_{z}^{2}\left|\psi_{1}^{+}\right\rangle}{\left\langle\psi_{1}^{+} \mid \psi_{1}^{+}\right\rangle}=1, \quad \frac{\left\langle\psi_{1}^{-}\right| \hat{S}_{z}^{2}\left|\psi_{1}^{-}\right\rangle}{\left\langle\psi_{1}^{-} \mid \psi_{1}^{-}\right\rangle}=1. 
\end{equation}
These equal 1 while the expectation values of all three Cartesian components squared and combined must give $\langle {\bf S}^2 \rangle =S(S+1)=6$. Considering further, these eigenstates of $\hat{H}$ are also eigenstates of $\hat{S}_{x}^{2}-\hat{S}_{y}^{2}$, with eigenvalues of ±3, which explains the energy eigenvalues.

Consider the even subspace with $m=0, \pm 2$ and the raising and lowering operations,
\begin{align}
\left(\hat{S}^{+}\right)^{2}|2\rangle & =0, & & \left(\hat{S}^{-}\right)^{2}|2\rangle=\sqrt{6-2} \sqrt{6}|0\rangle=\sqrt{24}|0\rangle, \nonumber \\
\left(\hat{S}^{+}\right)^{2}|-2\rangle & =\sqrt{6-2} \sqrt{6}|0\rangle=\sqrt{24}|0\rangle, & & \left(\hat{S}^{-}\right)^{2}|-2\rangle=0, \nonumber \\
\left(\hat{S}^{+}\right)^{2}|0\rangle & =\sqrt{6} \sqrt{6-2}|2\rangle=\sqrt{24}|2\rangle, & & \left(\hat{S}^{-}\right)^{2}|0\rangle=\sqrt{6} \sqrt{6-2}|-2\rangle=\sqrt{24}|-2\rangle . 
\end{align}
The operations with $\hat{S}_{z}^{2}$ on the $m$-states just give factors of $m^{2}$. Again let a potential eigenstate be written either as a ket superposition or as a column vector,
\begin{equation}
|\psi\rangle=a_{2}|2\rangle+a_{0}|0\rangle+a_{-2}|-2\rangle, \quad \text { or } \quad \psi=\left(\begin{array}{c}
a_{2}  \\
a_{0}  \\
a_{-2}
\end{array}\right) .
\end{equation}
When $\hat{H}$ acts on the states of $\hat{S}_{z}$, there results
\begin{align}
\hat{H}|2\rangle & =4 D|2\rangle+\frac{E}{2} \sqrt{24}|0\rangle, \nonumber \\
\hat{H}|0\rangle & =\frac{E}{2}(\sqrt{24}|2\rangle+\sqrt{24}|-2\rangle), \nonumber \\
\hat{H}|-2\rangle & =4 D|-2\rangle+\frac{E}{2} \sqrt{24}|0\rangle . 
\end{align}
Then the matrix for $\hat{H}$ in this subspace is constructed, producing the eigenvalue problem,
\begin{equation}
(\hat{H})(\psi)=\left(\begin{array}{ccc}
4 D & \sqrt{6} E & 0   \\
\sqrt{6} E & 0 & \sqrt{6} E \\
0 & \sqrt{6} E & 4 D
\end{array}\right)\left(\begin{array}{c}
a_{2} \\
a_{0} \\
a_{-2}
\end{array}\right)=\lambda\left(\begin{array}{c}
a_{2} \\
a_{0} \\
a_{-2}
\end{array}\right)=\lambda(\psi) .
\end{equation}
The determinant of $\hat{H}-\lambda \hat{I}$ must be zero for the solutions,
\begin{alignat}{1}
\operatorname{det} =\left|\begin{array}{ccc}
4 D-\lambda & \sqrt{6} E & 0 \\
\sqrt{6} E & -\lambda & \sqrt{6} E \\
0 & \sqrt{6} E & 4 D-\lambda
\end{array}\right| & =(4 D-\lambda)\left[\lambda(\lambda-4 D)-6 E^{2}\right]-(\sqrt{6} E)^{2}(4 D-\lambda)
\nonumber \\
& =(4 D-\lambda)\left(\lambda^{2}-4 D \lambda-12 E^{2}\right)=0 . 
\end{alignat}
The first eigenvalue is denoted with index 2, because it is composed from $|m|=2$ components, and the value is
\begin{equation}
\lambda_{2}=4 D . 
\end{equation}
The eigenvector comes from
\begin{equation}
(\hat{H}-\lambda \hat{I})(\psi)=\left(\begin{array}{ccc}
0 & \sqrt{6} E & 0 \\
\sqrt{6} E & -4 D & \sqrt{6} E \\
0 & \sqrt{6} E & 0
\end{array}\right)\left(\begin{array}{c}
a_{2} \\
a_{0} \\
a_{-2}
\end{array}\right)=\left(\begin{array}{l}
0 \\
0 \\
0
\end{array}\right) .
\end{equation}
This gives $a_{0}=0$ and $a_{2}=-a_{-2}$, so an unnormalized eigenvector is
\begin{equation}
\left|\psi_{2}\right\rangle=|2\rangle-|-2\rangle .
\end{equation}
This antisymmetric superposition is later found to be the second lowest energy eigenstate.

The last two eigenvalues that result from solving the remaining quadratic equation are
\begin{equation}
\lambda_{20}^{ \pm}=\frac{1}{2}\left[4 D \pm \sqrt{(4 D)^{2}+4\left(12 E^{2}\right)}\right]=2 D \pm 2 \sqrt{D^{2}+3 E^{2}}. 
\end{equation}
These are expressed concisely using a combined anisotropy parameter $D'$,
\begin{equation}
\label{eigs_1}
\lambda_{20}^{ \pm}= 2(D \pm D'), \quad D'\equiv \sqrt{D^2+3E^2}.
\end{equation}
These happen to be the highest and lowest eigenvalues, whose difference gives the span of the energy eigenvalues,
\begin{equation} \Delta_{\text{span}}=\lambda_{20}^{+}-\lambda_{20}^{-}= 4D' = 4\sqrt{D^2+3E^2}. \end{equation}
These states combine $|m|=2$ with $m=0$ components. The eigenvectors come from
\begin{equation}
\left(\hat{H}-\lambda_{20}^{ \pm} \hat{I}\right)(\psi)=\left(\begin{array}{ccc}
2 (D \mp D') & \sqrt{6} E & 0  \\
\sqrt{6} E & -2 (D \pm D') & \sqrt{6} E \\
0 & \sqrt{6} E & 2 (D \mp D')
\end{array}\right)\left(\begin{array}{c}
a_{2} \\
a_{0} \\
a_{-2}
\end{array}\right)=\left(\begin{array}{l}
0 \\
0 \\
0
\end{array}\right) .
\end{equation}
These constrain the amplitudes by
\begin{equation}
(2 D \mp 2 \alpha) a_{2}+\sqrt{6} E a_{0}=0 \quad \Longrightarrow \quad a_{0}=\frac{2(-D \pm D')}{\sqrt{6} E} a_{2}, \quad a_{-2}=a_{2} . 
\end{equation}
Then, two unnormalized eigenvectors are
\begin{equation}
\left|\psi_{20}^{ \pm}\right\rangle=|2\rangle+\sqrt{\frac{2}{3}} \frac{(-D \pm D')}{E}|0\rangle+|-2\rangle . 
\end{equation}
For these, it is helpful to use the fact that $D<0$ and $E \ll|D|$, so that the combinations $-D \pm D'$ can be approximated. One gets, by expanding $D'$ with the binomial theorem,
\begin{equation}
D'=|D| \sqrt{1+3 E^{2} / D^{2}} \approx|D|+\frac{3 E^{2}}{2|D|} . 
\end{equation}
This results in
\begin{equation}
-D+D' \approx 2|D|+\frac{3 E^{2}}{2|D|}=\text { large, positive, } \quad-D-D' \approx-\frac{3 E^{2}}{2|D|}=\text { small, negative. } 
\end{equation}
The corresponding energy eigenvalues are approximated as
\begin{equation}
\lambda_{20}^{+}=2(D+D') \approx \frac{3 E^{2}}{|D|}, \quad \lambda_{20}^{-}=2(D-D') \approx-4|D|-\frac{3 E^{2}}{|D|} . 
\end{equation}
The associated eigenvectors are approximated as
\begin{equation}
\left|\psi_{20}^{+}\right\rangle \approx|2\rangle+\sqrt{\frac{8}{3}} \frac{|D|}{E}|0\rangle+|-2\rangle, \quad \quad\left|\psi_{20}^{-}\right\rangle \approx|2\rangle-\sqrt{\frac{3}{2}} \frac{E}{|D|}|0\rangle+|-2\rangle, 
\end{equation}
The $|\psi_{20}^{+}\rangle$ state has a very large and positive $|0\rangle$ component, while the $|\psi_{20}^{-}\rangle$ state has a very small and negative $|0\rangle$ component.

The above describes the full energy eigenspectrum. The eigenvectors hold information about how the states are physically distinguished. One can verify that all the eigenvectors are mutually orthogonal to each other, as required.

\subsection{Eigenspectrum for Fe$^{2+}$:forsterite}
Let's check the eigenvalues/eigenvectors for Fe$^{2+}$:forsterite. With $D<0$ and $E \ll|D|$, the order from highest to lowest is as follows, calculated for $D=-12.37\ \mathrm{K}, E=0.896\ \mathrm{K}$, and $D'=12.46697\ \mathrm{K}$:
\begin{align}
\label{S2evs}
\lambda_{20}^{+} & = 2(D+D') \approx 0.19 \mathrm{~K}, & \left|\psi_{20}^{+}\right\rangle \approx|2\rangle+\sqrt{\frac{8}{3}} \frac{|D|}{E}|0\rangle+|-2\rangle, \nonumber \\
\lambda_{1}^{+} & =-|D|+3 E \approx-9.68 \mathrm{~K}, & \left|\psi_{1}^{+}\right\rangle=|1\rangle+|-1\rangle, \hskip 2.5cm \nonumber \\
\lambda_{1}^{-} & =-|D|-3 E \approx-15.06 \mathrm{~K}, & \left|\psi_{1}^{-}\right\rangle=|1\rangle-|-1\rangle, \hskip 2.5cm \nonumber \\
\lambda_{2} & =-4|D| \approx-49.48 \mathrm{~K}, &  \left|\psi_{2}\right\rangle=|2\rangle-|-2\rangle, \hskip 2.6cm \nonumber \\
\lambda_{20}^{-} &= 2(D-D') \approx-49.67 \mathrm{~K} , & \left|\psi_{20}^{-}\right\rangle \approx|2\rangle-\sqrt{\frac{3}{2}} \frac{E}{D \mid}|0\rangle+|-2\rangle,
\end{align}
The lowest state is dominated by its symmetric superposition of $m= \pm 2$ components, because the numerical factor on the $m=0$ ket is $-\sqrt{1.5} E /|D| \approx-0.0887$. The second lowest state has no dependence on $E$ or $D$, and it is simply the antisymmetric superposition of the $m= \pm 2$ states. The highest state, on the other hand, is dominated by the $m=0$ state, because $\sqrt{8 / 3}|D| / E \approx 22.5$, with a small symmetric contribution from the $m= \pm 2$ states. Therefore, the two lowest states are distinguished mainly by being symmetric and antisymmetric in the $m= \pm 2$ kets.

The gap $\Delta$ is the difference between the lowest two states, which are slightly split due to the $E$-interaction, in full and approximate expressions,
\begin{equation}
\Delta = \lambda_2-\lambda_{20}^{-} = 0.1939\ \mathrm{K},
\quad \text{or, } \quad
\Delta \approx \frac{3 E^{2}}{|D|} \approx 0.1947 \mathrm{~K} .
\end{equation}
Including Boltzmann's and Planck's constants, the zero-field gap corresponds to a microwave frequency,
\begin{equation}
\label{tunnelf}
f = k_{\rm B} \Delta /h =  4.040\ \text{GHz},
\end{equation}
which is the frequency at which the amplitudes of the primary components of the lowest two states, $|+2\rangle$ and $|-2\rangle$, oscillate if the system is initialized in one of them.
The intrinsic rate at which tunneling takes place between the $m= \pm 2$ components occurs at twice this frequency.  
%

The energy span between the highest and lowest levels is the difference of the extreme eigenvalues
\begin{equation}
\Delta_{\text{span}}=\lambda_{20}^{+}-\lambda_{20}^{-}= 49.8678\ \mathrm{K}, \quad \text{or, } \ \Delta_{\text{span}} \approx 4|D|+\frac{6 E^{2}}{|D|} \approx 49.8694 \mathrm{~K}. 
\end{equation}

\subsection{Fe(H$_2$O)$_6^{2+}$ transition energies}
\label{subsec:validation_2_transition_energies}

The Fe(H$_2$O)$_6^{2+}$ ion also has spin $S = 2$. We can use the eigenvalues found above with the known axial and transverse anisotropy terms: $D$ = 17.19 K and $E$ = 0.947 K, with a slight difference in the calculation, given that $D$ is positive:

Beginning with Eq.\ (\ref{eigs_1})

\begin{equation}
\lambda_{20}^{ \pm}= 2(D \pm D'), \quad D'\equiv \sqrt{D^2+3E^2}.
\end{equation}

Given that $D>0$ and $E \ll D$

\begin{equation}
D'=D \sqrt{1+3 E^{2} / D^{2}} \approx D+\frac{3 E^{2}}{2|D|} . 
\end{equation}

This results in
\begin{equation}
D+D' \approx 2D+\frac{3 E^{2}}{2D}=\text { large, positive, } \quad D-D' \approx-\frac{3 E^{2}}{2D}=\text { small, negative. } 
\end{equation}
The corresponding energy eigenvalues are approximated as
\begin{equation}
\lambda_{20}^{+}=2(D+D') \approx 4D + \frac{ 3 E^{2}}{D}, \quad \lambda_{20}^{-}=2(D-D') \approx-\frac{3 E^{2}}{D} . 
\end{equation}

\begin{equation} 
\begin{aligned}
\lambda_{20}^{-} &\approx -\frac{3 E^{2}}{D} \approx-0.16 \ K\\
\lambda_{1}^{-} & =D-3 E \approx14.36 \ K\\
\lambda_{1}^{+} & =D+3 E \approx20.03 \ K\\
\lambda_{2} & =4D \approx68.77 \ K\\
\lambda_{20}^{+} & \approx 4D + \frac{3 E^{2}}{D} \approx 68.93 \ K.
\end{aligned}
\end{equation}
These are the energy states of the system and not the transitions, which are obtained by taking the difference between the non-ground state energies and the ground state energy. 

\begin{equation} 
\begin{aligned}
\lambda_{1}^{-} \ & - \ \lambda_{20}^{-} \ = 14.49 \ K \\
\lambda_{1}^{+} \ & - \ \ \lambda_{20}^{-} \  = 20.19 \ K \\
\lambda_{2} \ & - \ \lambda_{20}^{-} \  = 68.93 \ K \\
\lambda_{20}^{+} \ & - \ \lambda_{20}^{-} \  = 69.09 \ K . 
\end{aligned}
\end{equation}

\section{$\mathbf{S=2}$ Diagonalization with Zeeman term}
\label{app:S=2_analytic_solution_zeeman}

The quantum spin Hamiltonian with the Zeeman term is: 
\begin{equation}
    \hat{H} = D\hat{S}_z^2 + E(\hat{S}_x^2 - \hat{S}_y^2) + g_L \mu_B B_z \hat{S}_z
    \label{eq:hamiltonian_static_2}
\end{equation}
where $g_{L} \approx 2$ is the Landé $g$-factor, and $\mu_{B}=e \hbar /\left(2 m_{\mathrm{e}}\right)$ is the electron's Bohr magneton. The last term is the Zeeman interaction, $U_{\mathrm{Z}}=-\boldsymbol{\mu} \cdot \mathbf{B}$. The sign on the Zeeman interaction in $\hat{H}$ is positive because the magnetic dipole moment is opposite to $\mathbf{S}$ due to the negative charge of the electrons, as in $\boldsymbol{\mu}=-g_{L} \mu_{B} \mathbf{S}$. 

In order to calculate the eigenvalues of the two subspaces: $m = \pm 1$ and $m = 0, \pm 2$, we follow the same steps shown in Appendix~\ref{app:S=2_analytic_solution}, while incorporating the effect of the Zeeman term in the matrices, which is proportional to $z\equiv g_L\mu_B B_z$. 
For the m = $\pm$ 1 subspace:
\begin{align}
\langle +1|\hat H|+1\rangle &= D(1)^2 + z(1) = D + z \nonumber \\
\langle -1|\hat H|-1\rangle &= D(-1)^2 + z(-1) = D - z \nonumber \\
\langle +1|\hat H|-1\rangle &= \langle -1|\hat H|+1\rangle = 3E
\end{align}
The Zeeman term appears only on the diagonal of the matrix:
\begin{equation}
(\hat{H})=\left(\begin{array}{cc}
D + z & 3 E  \\
3 E & D - z
\end{array}\right).
\end{equation}
Solve the eigenvalue problem: 
\begin{equation}
(\hat{H})(\psi)=\left(\begin{array}{cc}
D + z & 3 E  \\
3 E & D - z
\end{array}\right)\binom{a_{1}}{a_{-1}}=\lambda\binom{a_{1}}{a_{-1}}=\lambda(\psi) .
\end{equation}
The solutions are found from the determinant of $\hat{H}-\lambda \hat{I}$ being zero,
\begin{equation}
\left|\begin{array}{cc}
D + z - \lambda & 3 E  \\
3 E & D-z-\lambda
\end{array}\right|=(D-\lambda)^{2}-z^2 -9 E^{2}=0 \quad \Longrightarrow \lambda_{1}^{\pm}= D \pm \sqrt{9E^2+z^2} .
\end{equation}
In the main text, an applied field of $5.0$ T is considered for Fe$^{2+}$:forsterite. 
Using the appropriate values, $D = -12.37$ K, $E = 0.896$ K and $z = g_L \mu_B B_z \approx 6.717$, results in the eigenvalues being:
\begin{equation}
\lambda_\pm = D \pm \sqrt{9E^2+z^2} = -12.37 \pm \sqrt{9(0.896)^2+(6.717)^2} = 
\begin{cases} \lambda_1^{+} = -5.135\ \mathrm{K}, \\ 
\lambda_1^{-} = -19.605\ \mathrm{K}.  \end{cases} \end{equation}
%
%
We are not showing the calculation of the eigenvectors, as they are  not essential for the discussion in the main text. 

For the subspace with $m = 0, \pm 2$, there are matrix elements
\begin{align}
\langle +2|\hat H|+2\rangle &= D(2)^2 + z(2) = 4D + 2z \nonumber \\
\langle 0|\hat H|0\rangle &= D(0)^2 + z(0) = 0 \nonumber \\
\langle -2|\hat H|-2\rangle &= D(-2)^2 + z(-2) = 4D - 2z \nonumber \\
\langle +2|\hat H|0\rangle &= \langle 0|\hat H|+2\rangle = \sqrt6\,E \nonumber \\
\langle 0|\hat H|-2\rangle &= \langle -2|\hat H|0\rangle = \sqrt6\,E \nonumber \\
\langle +2|\hat H|-2\rangle &= \langle -2|\hat H|+2\rangle = 0
\end{align}
which give the Hamiltonian matrix
\begin{equation}
\hat{H} = \begin{pmatrix} 4D+2z & \sqrt6\,E & 0 \\ \sqrt6\,E & 0 & \sqrt6\,E \\ 0 & \sqrt6\,E & 4D-2z \end{pmatrix}.
\end{equation}
The determinant of $\hat{H}-\lambda \hat{I}$ must be zero for the solutions,
\begin{align}
\det\left(H_{even}-\lambda I\right) &= \det\begin{pmatrix} 4D+2z-\lambda & \sqrt6\,E & 0 \\ \sqrt6\,E & -\lambda & \sqrt6\,E \\ 0 & \sqrt6\,E & 4D-2z-\lambda \end{pmatrix} = 0 \\[2mm] \nonumber 
&= -\lambda^3 + 8D\lambda^2 + (12E^2+4z^2-16D^2)\lambda - 48DE^2 = 0
\end{align}
For $B_z=5.0$ T and the values $D = -12.37$ K, $E = 0.896$ K and $z = g_L \mu_B B_z \approx 6.717$ K, the equation is 
\begin{equation}
\lambda^3 + 98.96\,\lambda^2 + 2258.16\,\lambda - 476.68 = 0
\end{equation}
The roots are: 
\begin{equation}
\lambda_{20}^{-} \approx -62.991\ \text{K}, \qquad \lambda_{2} \approx -36.178\ \text{K}, \qquad \lambda_{20}^{+} \approx 0.209\ \text{K}.
\end{equation}
These have been identified with the corresponding eigenvalues for $z=0$ by tracking their changes with increasing $z$.
From highest to lowest, the eigenvalues of the system are
\begin{align}
\lambda_{20}^{+} &\approx 0.209\ \text{K}, \nonumber \\
\lambda_{1}^{+} &= -5.135\ \text{K}, \nonumber \\
\lambda_{1}^{-} &= -19.605\ \text{K}, \nonumber \\
\lambda_{2} &\approx -36.178\ \text{K}, \nonumber \\
\lambda_{20}^{-} &\approx -62.991\ \text{K}.
\end{align}
Due to the strong magnetic field, the lowest eigenstate is dominated by $m=-2$, i.e., $|\psi_{20}^{-}\rangle \approx |-2\rangle$, while the second lowest eigenstate is dominated by $m=+2$, i.e., $|\psi_2\rangle \approx |2\rangle$.  Both have a small admixture with the opposite sign of $m$ and with $m=0$. 
The energy gap between the two lowest states is
\begin{equation}
\Delta = \lambda_{2} - \lambda_{20}^{-} = -36.178 - (-62.991) = 26.81\ \text{K}.
\end{equation}
Tunnel splitting cannot occur with the applied field of 5.0 T because the energy gap of  26.81 K is much larger than the coupling that mixes them, $\sqrt{6}E$ $\approx$ 2.2 K. Indeed, ignoring $E \ll z$ in the Hamiltonian matrix, it becomes diagonal with its strong-field eigenvalues well-approximated by $\lambda = 0, 4D \pm z, 4D \pm 2z$, producing a gap $\Delta \approx 4z \approx 26.868$ K.


\bibliography{aipsamp}

\end{document}